\documentclass{article}

\usepackage[preprint,nonatbib]{neurips_2020}
\usepackage[toc,page,titletoc]{appendix}

\usepackage[utf8]{inputenc} 
\usepackage[T1]{fontenc}    
\usepackage{hyperref}       
\usepackage{url}            
\usepackage{booktabs}       
\usepackage{amsfonts}       
\usepackage{nicefrac}       
\usepackage{microtype}      
\usepackage{amsmath}

\usepackage{graphicx}
\usepackage{float} 
\usepackage{subcaption}

\DeclareMathAlphabet\mathbfcal{OMS}{cmsy}{b}{n} 

\title{Predictive coding in balanced neural networks with noise, chaos and delays}

\author{
Jonathan Kadmon\thanks{Correspondence to kadmonj@stanford.edu}\\
  Department of Applied Physics\\
  Stanford University, CA\\
  \And
  Jonathan Timcheck\\
  Department of Physics\\
  Stanford University, CA \\
  \And
  Surya Ganguli\\
  Department of Applied Physics\\
  Stanford University, CA\\
  }

\begin{document}

\maketitle

\begin{abstract}

Biological neural networks face a formidable task: performing reliable computations in the face of intrinsic stochasticity in individual neurons, imprecisely specified synaptic connectivity, and nonnegligible delays in synaptic transmission.  A common approach to combatting such biological heterogeneity involves averaging over large redundant networks of $N$ neurons resulting in coding errors that decrease classically as $1/\sqrt{N}$. Recent work demonstrated a novel mechanism whereby recurrent spiking networks could efficiently encode dynamic stimuli,  achieving a superclassical scaling in which coding errors decrease as $1/N$. This specific mechanism involved two key ideas: predictive coding, and a tight balance, or cancellation between strong feedforward inputs and strong recurrent feedback.  However, the theoretical principles governing the efficacy of balanced predictive coding and its robustness to noise, synaptic weight heterogeneity and communication delays remain poorly understood.  To discover such principles, we introduce an analytically tractable model of balanced predictive coding, in which the degree of balance and the degree of weight disorder can be dissociated unlike in previous balanced network models, and we develop a mean field theory of coding accuracy. Overall, our work provides and solves a general theoretical framework for dissecting the differential contributions neural noise, synaptic disorder, chaos, synaptic delays, and balance to the fidelity of predictive neural codes, reveals the fundamental role that balance plays in achieving superclassical scaling, and unifies previously disparate models in theoretical neuroscience. 

\end{abstract}

\section{Introduction}

The early days of computing generated intense interest in how reliable computations could emerge from unreliable components, a question well articulated by von Neumann \cite{Von_Neumann1956-ze}.  While the rise of digital technology largely circumvented this issue by making individual physical components highly reliable and fast, biological evolution, in the case of neural computation, had to directly face this problem.  Indeed neural cortical firing patterns exhibit high levels of Poisson like temporal irregularity \cite{abeles1991corticonics,softky1993highly,Shadlen1998-wk}, external noisy inputs to a circuit can interfere with its operation, synaptic strengths are imprecisely specified in learning and development \cite{chambers2017stable}, and synapses themselves can be slow \cite{greengard2001neurobiology}, resulting in non-negligible communication delays between neurons.  Thus von Neumann's question still remains central in neuroscience: how can neural circuits perform reliable computations when their underlying components, connectivity and inputs can be slow and subject to unpredictable fluctuations?    

A conventional approach to this problem involves averaging over large redundant networks of $N$ neurons, resulting in coding and computation errors that decay as $O(1/\sqrt{N})$ as long as neural firing patterns are weakly correlated, due to the law of large numbers.  However, can one do better? Recent work \cite{Boerlin2013-ks} has constructed a recurrent network of spiking neurons that achieves \textit{superclassical} error scaling, with the error decreasing as $O(1/N)$.  Two key ideas underlying this network are the notions of predictive coding \cite{rao1999predictive,eliasmith2004neural} and balance \cite{Deneve2016-jv}. In a sensory coding context, predictive coding refers to scenarios in which a neural circuit computes a prediction $\hat x(t)$ of some dynamic sensory input $x(t)$. Then a representation of the prediction error $\hat x(t) - x(t)$ can be employed for diverse purposes, including learning a causal model \cite{giret2014evidence}, cancellation of predictable sensory consequences of motor actions \cite{keller2012sensorimotor}, mismatch between auditory and visual speech perception \cite{arnal2011transitions}, or simply communicating surprises to downstream regions \cite{keller2018predictive}.  In \cite{Boerlin2013-ks} in particular, the prediction error was used to drive the dynamics of the recurrent spiking network through extremely strong negative feedback, thereby forcing the network prediction $\hat x(t)$ to track the sensory input $x(t)$.  Furthermore,  the gain $b$ of the negative feedback was proportional to network size $N$, resulting in a very  \emph{tight balance} or cancellation between strong feedforward drive due to the external input $b x(t)$ and recurrent negative feedback generated by the network prediction $- b \hat x(t)$.  

A notion of balance has also played a prominent role in theoretical neuroscience in the context of a very different question:  what mechanisms can generate the strong heterogeneity of observed biological firing patterns \cite{abeles1991corticonics,softky1993highly} in the first place? \cite{sompolinsky1988chaos,wolf2014dynamical} demonstrated that disordered random connectivity itself can generate fluctuations in firing activity due to high dimensional chaos in neural circuits, without the need for additional injected noise.  Moreover, recurrent networks in which each neuron receives strong excitation and strong inhibition, self-organize into a highly heterogenous balanced state \cite{Van_Vreeswijk1996-rm,Landau2018-jf}, where excitation and inhibition into each neuron is large and $O(\sqrt{N})$, but their difference cancels to $O(1)$ fluctuations which drive firing, a situation we term \emph{classical balance}, in contrast to the \emph{tight balance} of \cite{Boerlin2013-ks}.  Given the empirically observed prevalence of highly heterogenous firing patterns in the brain, the dynamical operating regime of cortex, and in particular, the degree of excitation-inhibition balance involved (tight, classical, or something looser) remains a question of great interest \cite{Hennequin2018-uo,ahmadian2019dynamical}.

These two largely distinct strands of inquiry, namely exploiting tight balance to make predictive coding highly efficient, versus exploiting classical balance to explain the origins of neural variability itself, in the absence of any particular computations, raises several foundational questions.  First, what is the relation between the chaotic networks of classical balance and the predictive coding networks of tight balance?  What minimal degree of balance can generate superclassical scaling of error with network size? Indeed can we elucidate the fundamental role of balance in achieving superclassical scaling?  Moreover, what is the efficacy of balanced predictive coding in the presence of noisy external inputs, chaos induced by additional weight disorder, or delays due to slow synaptic communication?  While some of the latter issues have been explored numerically in predictive coding spiking networks \cite{Schwemmer2015-jp,Chalk2016-zr}, a theoretical analysis of the interplay between balance, weight disorder, noise, chaos and delays in determining the fidelity of predictive coding has remained elusive due to the complexity of the network models involved.  This lack of understanding of how multiple facets of biological variablity interact with each other in predictive coding represents a major gap in the theoretical literature, given the prevalence of predictive coding in many areas of theoretical neuroscience \cite{rao1999predictive,eliasmith2004neural}.

We aim to fill this gap by introducing and analyzing a theoretically tractable neural network model of balanced predictive coding.  Importantly, in our new model we can independently adjust the amounts of: balance employed in predictive coding, weight disorder leading to chaos, strength of noise, degree of delay, and the single neuron nonlinearity. In previous balanced network models for generating heterogeneity, the degree of chaos inducing weight disorder and the degree of excitation-inhibition balance were inextricably intertwined in the same random connectivity pattern \cite{Van_Vreeswijk1996-rm}.  Our model in contrast exhibits an interplay between low rank structured connectivity implementing balance, and high rank disordered connectivity inducing chaos, each with {\it independently} adjustable strengths.  In general, how computation emerges from an interplay between structured and random connectivity has been a subject of recent interest in theoretical neuroscience \cite{Landau2018-jf,Mastrogiuseppe2018-sv,Mastrogiuseppe2018-hz,Rivkind2017-fp}.  Here we show how structure and randomness interact by obtaining analytic insights into the efficacy of predictive coding, dissecting the individual contributions of balance, noise, weight disorder, chaos, delays and nonlinearity, in a model were all ingredients can coexist and be independently adjusted.

\section{Linearly decodable neural codes in noisy nonlinear recurrent networks}
\label{sec:ErrorScale}
Consider a noisy, nonlinear recurrent neural network of $N$ neurons with a dynamical firing rate vector given by $\mathbf{r}(t)\in\mathbb{R}^N$.  We wish to encode a scalar dynamical variable $x(t)$ within the firing rate vector $\mathbf{r}(t)$ such that it can be read out at any time $t$ by a simple {\it linear} decoder $\hat{x}(t) =  \frac{1}{N}\mathbf{w}^T\mathbf{r}(t)$ where $\mathbf{w}$ is a {\it fixed} time-independent readout vector.  The dynamical variable $x(t)$ could be thought of either as an input stimulus provided to the network, or as an efferent motor command generated internally by the network as an autonomous dynamical system \cite{Alemi2018-cj}.  For simplicity, in the main paper we focus on the case of stimulus encoding, and describe how our analysis can be generalized to autonomous signal generation in the Supplementary Material (SM) in a manner similar to previous studies of efficient coding of dynamical systems in spiking networks \cite{Boerlin2013-ks,Alemi2018-cj,Kushnir2019LearningTS}.  Also, while we focus on scalar stimuli in the main paper, our theory can be easily generalized to multidimensional stimuli (see SM).

We assume the nonlinear dynamics of the firing rate vector $\mathbf{r}(t)$ obeys standard circuit equations \cite{amari1972characteristics}
\begin{equation}
\label{eq:circuit}
r_i(t) = \phi(h_i(t)),  \qquad \text{and} \qquad \tau\dot{h}_i(t)=-h_i(t)+\sum_jJ_{ij} r_j(t-d) + I_i(x(t)) + \sigma\xi_i(t). 
\end{equation}
Here $h_i(t)$ is the membrane potential of neuron $i$, $\phi $ is a neural nonlinearity that converts membrane potentials $h_i$ to output firing rates $r_i$, $\tau$ is the membrane time constant,  $J_{ij}$ is the synaptic connectivity from neuron $j$ to $i$, $d$ is a synaptic communication delay, $I_i(x(t))$ is the stimulus driven input current to neuron $i$, and $\xi_i(t)$ is zero mean i.i.d Gaussian white noise current input with cross-correlation $\langle \xi_i(t)\xi_j(t') \rangle=\delta_{ij}\delta(t-t')$. Now the critical issue is, how do we choose the connectivity $J_{ij}$ and the stimulus driven current $I_i(x)$ so that the noisy nonlinear delay dynamics in \eqref{eq:circuit} for $r_i(t)$ yields a simple linearly decodable neural code with a network estimate $\hat{x}(t) =  \frac{1}{N} \sum_i w_i r_i(t)$ closely tracking the true stimulus $x(t)$?  We generalize a proposal made in \cite{Boerlin2013-ks}, that was proven to be optimal in the case of spiking neural networks with no delays, noise or weight disorder, by choosing 
\begin{equation}
\label{eq:connectivity}
    J_{ij} =  g \mathcal J_{ij}  -  \frac{b}{N} w_i w_j,  \qquad \text{and} \qquad   I_i(x(t)) = b w_i x(t).
\end{equation}
Here, $w_i$ are the components of the readout vector, which now appear both in the stimulus driven current $I_i$ and the connectivity $J_{ij}$ in a structured rank $1$ manner.  We also consider a random contribution $g \mathcal J_{ij}$ to synaptic strengths,  modelling imprecision in connectivity.  We take the structured connectivity to be random with $w_i$ chosen i.i.d from a distribution $\mathcal P(w)$ such that $w_i$ remains $O(1)$ for large $N$ with the norm of the vector concentrating at $\mathbf{w}^T\mathbf{w} = N$, while the random synaptic strengths $\mathcal J_{ij}$ are chosen to be i.i.d Gaussian variables with zero mean and variance $\frac{1}{N}$.  Thus while the structured connectivity, which is $O(1/N)$, is much weaker than the random connectivity, which is $O(1/\sqrt{N})$, they each generate a comparable $O(1)$ contribution to the input current to any neuron (when $b$ is $O(1)$). Thus in this model, as $N \rightarrow \infty$, the input current to each neuron originates from $4$ distinct sources, with $3$ independently adjustable control strengths:  input currents due to disordered connectivity ($g$), structured connectivity ($-b$), stimulus drive ($+b$), and noise ($\sigma$).  

\begin{figure}

\centering
\includegraphics[width=\textwidth]{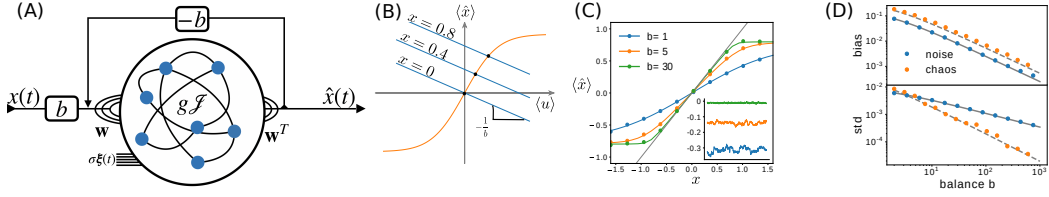} 
  \caption{\textbf{(A)}  A schematic view of a balanced predictive coding network. 
  \textbf{(B)} Graphical solution method for mean field equations in \eqref{eq:firstorder} for $\phi=\tanh$. 
  \textbf{(C)} The mean input-output transfer function $\langle \hat x \rangle$ as a function of $x$ obtained by solving \eqref{eq:firstorder} (solid curves) and numerical simulations of \eqref{eq:predcode} (points) with $N=1400$, $\sigma=0.75$ and $g=d=0$ for $3$ values of $b$. Grey line marks $\langle\hat{x}\rangle=x$. The inset shows $3$ corresponding examples of traces of $\hat x(t)-x$ when $x=0.5$, demonstrating both bias ($y$-axis baseline) and fluctuations $\delta \hat x(t)$. 
  \textbf{(D)} The decoder bias $\langle | x - \langle \hat x \rangle | \rangle$ (top) and standard deviation $\sqrt{\langle (\delta \hat x)^2 \rangle}$ (bottom) as a function of balance $b$ for theory (curves) and simulations (points). $\sigma=0.75$, $g=0$ for noise (blue), $\sigma=0$, $g=1.6$ for chaos (orange). In both cases $N=1400$ and $x=0.2$. Balance $b$ yields power law suppression of variance with exponent $-1$ for noise and $-2$ for chaos. \label{fig1}}
\end{figure}

Interestingly, this model provides a simple and theoretically tractable instantiation of the principle of predictive coding of the stimulus through balance (See Fig. \ref{fig1}A).  One can see this by inserting the connectivity in \eqref{eq:connectivity} into \eqref{eq:circuit} and using the definition of the readout $\hat{x}(t) =  \frac{1}{N} \sum_j w_j r_j(t)$ to obtain
\begin{equation}
\label{eq:predcode}
    \tau\dot{h}_i(t)= -h_i(t) + \sum_j g \mathcal{J}_{ij} r_j(t-d) + b w_i \left[ x(t) - \hat x(t-d) \right] + \sigma\xi_i(t).
\end{equation}
Thus the structured part of the recurrent connectivity implicitly computes a prediction of the stimulus $\hat x(t-d)$, which is then used to cancel the actual incoming stimulus $x(t)$, and the resulting coding error $x(t) - \hat x(t-d)$ drives membrane voltages $h_i$ in the readout direction $w_i$.  The coefficient $b$ defines a level balance between positive feedforward stimulus drive, and negative feedback from the prediction computed by the structured connectivity.  A key feature of our model is that, unlike in previous balanced network models \cite{Van_Vreeswijk1996-rm,Landau2018-jf,Renart2007-kz}, the degree of balance $b$ can be independently modulated relative to the degree of synaptic disorder, which here is controlled instead by $g$.  Moreover, through different choices of scaling of $b$ with $N$, we can seamlessly interpolate between previously distinct regimes of balance, with $b=O(N)$ corresponding to tight balance \cite{Deneve2016-jv}, $b=O(\sqrt{N})$ corresponding to classical balance \cite{Landau2018-jf}, and $b<O(\sqrt{N})$ corresponding to loose or no balance \cite{ahmadian2019dynamical,Mastrogiuseppe2018-sv,rubin2015stabilized}.

However, despite the prominent role of both balanced networks (e.g., \cite{Renart2007-kz,Roudi2007-kn,Monteforte2010-hq,Rosenbaum2014-mi,Pattadkal2018-vq}) and predictive coding \cite{rao1999predictive,keller2012sensorimotor,arnal2011transitions,keller2018predictive} in theoretical neuroscience, to our knowledge, an analytic theory of the robustness of balanced predictive coding in the face of weight disorder, noise and delays in general nonlinear networks has not yet been developed.  We take advantage of our simple analytically tractable model of balanced predictive coding in \eqref{eq:circuit} and \eqref{eq:connectivity} to compute how the average error $\varepsilon^2 =  \langle \left [x(t) - \hat x(t) \right ]^2 \rangle$ of the neural code depends on various network properties.  We work in an adiabatic limit in which the external stimulus $x(t)$ varies over a much longer time scale $T$ than either the membrane time constant $\tau$ or the delay $d$.  Thus we can think of the stimulus $x(t)$ as effectively a constant $x$, and the squared error arises as the sum of a squared bias and a variance: $\varepsilon^2 = (\langle \hat x \rangle - x)^2  +  \langle (\delta \hat x)^2 \rangle$, where $\delta \hat x = \hat x - \langle \hat x \rangle$.  The average $\langle \cdot \rangle$ can be thought of as an average over the realizations of noise $\xi_i$, or equivalently, a temporal average over an intermediate window of duration between that of the microscopic times scales of $\tau$ and $d$ and the macroscopic time scale $T$.  Our goal in the following is to compute the bias and variance by computing the mean and variance of $\hat x(t)$ and its dependence on the strengths of noise $\sigma$, balance $b$, weight disorder $g$, delay $d$, and nonlinearity $\phi$.

\section{A mean-field theory for bias and variance in a noisy neural code}
\label{sec:noise}

We first consider the case of no weight disorder and delay ($g=d=0$ in \eqref{eq:predcode}), focusing on the interplay between balance $b$, nonlinearity $\phi$ and noise strength $\sigma$.  To analyze these dynamics, we first decompose the membrane voltage vector $\mathbf h$ into two components, parallel and perpendicular to the readout vector $\mathbf w$, via $\mathbf h(t) = \mathbf h^{\parallel}(t) + \mathbf h^{\perp}(t)$ where $\mathbf h^{\parallel} = \mathbfcal P \mathbf h$ and $\mathbf h^{\perp} = (\mathbf I - \mathbfcal P) \mathbf h$, and $\mathbfcal P = \frac{1}{N} \mathbf w \mathbf w^T$ is an orthogonal projection operator onto the direction of $\mathbf w$.  Thus $\mathbf h^{\parallel}(t) = u(t) \mathbf w$ where $u(t) \equiv \frac{1}{N} \sum_{i=1}^N w_i h_i(t)$ and $\mathbf h^{\perp}$ obeys $\mathbf w^T \mathbf h^{\perp} = 0$.  Now applying $\frac{1}{N} \mathbf w^T$  and $\mathbf I - \mathbfcal P$ to both sides of \eqref{eq:predcode} we can decompose the dynamics into that of $u(t)$ and $h_i^{\perp}(t)$ respectively:  
\begin{equation}
\label{eq:decompose}
\tau \dot u(t) = -u(t) + b \left[ x - \hat x(t) \right] + \sigma \xi^{\parallel}(t),  \qquad \text{and} \qquad
\tau \dot h^\perp_i(t)  =  - h^\perp_i(t) + \sigma \xi^{\perp}_i(t). 
\end{equation}
The noise $\xi^\parallel = \frac{1}{N} \sum_{i=1}^N w_i \xi_i$ 
along the decoder direction now has diminished autocorrelation $\langle \xi^\parallel(t)\xi^\parallel(t') \rangle=\frac{1}{N}\delta(t-t')$, 
while the perpendicular noise components have autocorrelation $\langle \xi^\perp_i(t)\xi^\perp_j(t') \rangle=\delta_{ij}\delta(t-t')$ 
up to $O(1/N)$ corrections due to satisfying the constraint $\sum_i w_i \xi^\perp_i = 0$, 
which we can safely neglect. Thus in the large $N$ limit, the variables $h^\perp_i(t)$ undergo independent 
Ornstein Uhlenbeck (OU) processes each corresponding to leaky integration with time constant $\tau$ of white noise of variance $\sigma^2$, yielding an output with zero mean and temporal variance $\langle (h^\perp_i)^2 \rangle = \frac{\sigma^2}{2\tau}$.  

Next, in order to compute the temporal mean and variance of $\hat x(t)$, we decompose $u(t)$ into its temporal mean $\langle u \rangle$ and fluctuations $\delta u(t)$ about that mean via $u(t) = \langle u \rangle + \delta u(t)$.  Inserting this decomposition into the dynamical equation for $u(t)$ in \eqref{eq:decompose} and taking the temporal average $\langle \cdot \rangle$ of both sides, we obtain the relation $\langle u \rangle = b \left[ x - \langle \hat x \rangle \right]$. We can obtain a second relation between $\langle \hat x \rangle$ and $\langle u \rangle$ by starting from the definition of $\hat x(t)$ and inserting the decompositions $h_i(t) = w_i u(t) + h^\perp_i(t)$ and $u(t) = \langle u \rangle + \delta u(t)$ to obtain $\hat x(t) = \frac{1}{N} \sum_{i=1}^N w_i \phi(h_i(t)) = \frac{1}{N} \sum_{i=1}^N w_i \phi \left(w_i \langle u \rangle + w_i \delta u(t) + h^\perp_i(t)\right)$. Now since $u(t)$ is driven by white noise $\xi^\parallel(t)$ of variance $O(1/N)$ in \eqref{eq:decompose}, we expect the fluctuations $\delta u(t)$ in the coding direction $\mathbf w$ to be of variance $O(1/N)$, and therefore much smaller than either the mean $\langle u \rangle$ or the perpendicular membrane voltages $h^\perp_i(t)$, both of $O(1)$, inside the argument of $\phi$.  Therefore we Taylor expand the nonlinearity $\phi$ about $\delta u(t) = 0$ to obtain to first order in $\delta u$:
\begin{equation}
\label{eq:linearhatx}
    \hat x(t) = \frac{1}{N} \sum_{i=1}^N w_i \phi \left(w_i \langle u \rangle + h^\perp_i(t)\right) + \frac{1}{N} \sum_{i=1}^N w^2_i \phi' \left(w_i \langle u \rangle + h^\perp_i(t)\right) \delta u(t).
\end{equation}
Now, taking the temporal average $\langle \cdot \rangle$ of both sides of this equation, we obtain, up to corrections of $O(\frac{1}{N})$, 
 $  \langle \hat x \rangle  = \frac{1}{N} \sum_{i=1}^N w_i \langle \phi \left(w_i \langle u \rangle + h^\perp_i(t)\right) \rangle 
    = \int \mathcal Dz \, dw \, \mathcal P(w)  w \phi(w \langle u \rangle + \frac{\sigma}{\sqrt{2\tau}} z)$.
Here $\mathcal P(w)$ is the distribution of readout weights and $\mathcal Dz = \frac{dz}{\sqrt{2\pi}}e^{-z^2/2}$ is the standard Gaussian measure.  Thus we have obtained two equations for the two unknown means $\langle \hat x \rangle$ and $\langle u \rangle$:
\begin{equation}
\label{eq:firstorder}
    \langle \hat x \rangle = x - \frac{\langle u \rangle}{b}, \qquad \text{and} \qquad \langle \hat x \rangle = \int \mathcal Dz \, dw \, \mathcal P(w)  w \phi(w \langle u \rangle + \frac{\sigma}{\sqrt{2\tau}} z).
\end{equation}
The solutions to these equations can be viewed graphically (Fig. \ref{fig1}B).  The first equation describes a straight line in the $\langle u\rangle$-$\langle\hat{x}\rangle$  plane with intercept $x$ and slope $-1/b$ (blue curves).  The second equation behaves like a smoothed version of the nonlinearity $\phi$ (orange curve), and the intersection of these curves yields the solution. Thus as $b$ is increased, the slope of the line flattens, and the bias $|\langle\hat{x}\rangle-x|$ decreases, as long as $x$ lies in the dynamical range of the smoothed $\phi$.  In general, the input-output behavior $x \rightarrow \langle \hat x \rangle$ is largely linear for all such values of $x$ at large $b$ (Fig. \ref{fig1}C).  Our quantitative predictions for the bias are confirmed via numerical simulations in Fig. \ref{fig1}D, top.  With knowledge of the nonlinearity $\phi$, degree of balance $b$, and noise level $\sigma$, one can theoretically compute the deterministic bias and remove it through the inverse map $\langle \hat x \rangle \rightarrow x$ when feasible.  Therefore, we focus on the contribution of variance $\langle [ \delta \hat x(t) ]^2 \rangle$ to coding error $\varepsilon$, which cannot be easily removed.

To compute the variance of $\delta \hat x$, we insert the decompositions $u(t) = \langle u \rangle + \delta u(t)$ and $\hat x(t) = \langle \hat x \rangle + \delta \hat x(t)$ into \eqref{eq:decompose} and use the mean relation $-\langle u \rangle + b \left[ x - \langle \hat x \rangle \right] = 0$ to extract a dynamic equation for the fluctuations $\tau \dot \delta u(t) = -\delta u(t) - b \delta \hat x(t) + \sigma \xi^{\parallel}(t)$.  We then subtract $\langle \hat x \rangle$ from both sides of \eqref{eq:linearhatx} to obtain the linearized relation $\delta \hat x(t) = \langle \phi' \rangle \delta u$ where $\langle \phi' \rangle \equiv \frac{1}{N} \sum_{i=1}^N w^2_i \phi' \left(w_i \langle u \rangle + h^\perp_i(t)\right).$   Inserting this relation into $\dot \delta u(t)$ and replacing the sum over $i$ with integrals yields 
\begin{equation}
\label{eq:phiprime}
    \tau \delta \dot u(t) = -\delta u(t) - b \langle \phi' \rangle \delta u(t) + \sigma \xi^{\parallel}(t) \quad \text{where} \quad 
    \langle \phi' \rangle = \int \mathcal Dz \, dw \, \mathcal P(w)  w^2 \phi'(w \langle u \rangle + \frac{\sigma}{\sqrt{2\tau}} z).
\end{equation}
This constitutes a dynamic mean field equation for the membrane voltage fluctuations $\delta u(t)$ in the coding direction $\mathbf w$, where the average gain of the nonlinearity $\langle \phi' \rangle$ across neurons multiplicatively  modifies the negative feedback due to balance $b$.  Again, this is an OU process like that of $h^\perp_i$ in \eqref{eq:decompose} except with a faster effective time constant $\tau_\text{eff} = \frac{\tau}{1+b \langle \phi' \rangle}$ and a smaller input noise variance $\sigma^2_\text{eff} = \frac{\sigma^2}{N(1+b \langle \phi' \rangle)^2}$  yielding a diminished variance $\langle (\delta u(t))^2 \rangle  = \frac{\sigma^2}{2\tau N(1+b \langle \phi' \rangle)}$ both due to effective negative feedback, and averaging over the decoder direction $\mathbf w$. Note the fluctuations of $\delta u$ are indeed $O(1/N)$ making our initial assumption self-consistent.  Finally, the variance of the readout fluctuations follows from squaring and averaging both sides of $\delta \hat x(t) = \langle \phi' \rangle \delta u(t)$, yielding
\begin{equation}
\label{eq:secondorder}
    \langle (\delta u(t))^2 \rangle  = \frac{\sigma^2}{2\tau N(1+b \langle \phi' \rangle)},  \qquad \text{and}  \qquad 
    \langle (\delta \hat x(t))^2 \rangle = \frac{\langle \phi' \rangle^2 \sigma^2}{2\tau N(1+b \langle \phi' \rangle)}.
\end{equation}
Taken together, the equations \eqref{eq:firstorder}, \eqref{eq:phiprime} and \eqref{eq:secondorder} constitute a complete mean field theory of how the first and second order statistics of the projection of the membrane voltages and firing rates onto the decoder direction $\mathbf w$, i.e. 
$u(t) = \frac{1}{N} \sum_i w_i h_i(t)$ and $\hat x(t) = \frac{1}{N} \sum_i w_i r_i(t)$ respectively, depend on the balance $b$ and noise $\sigma$, in the large $N$ limit. We compare the theoretically predicted decoder bias $\langle \hat x \rangle - x$ and variance $\langle (\delta \hat x)^2 \rangle$ with numerical experiments, obtaining an excellent match (see Fig. \ref{fig1}D and Figures below).  We find that the standard deviation of the decoder output scales as $O(1/b\sqrt N)$.  This reveals a fundamental necessity of strong balance, in which $b$ must scale as $N^\chi$ for some $\chi>0$, to achieve superclassical scaling with decoder error falling off faster than $O(1/\sqrt N)$.

\section{The interplay between balance and chaos induced by weight disorder}
\label{sec:balancechaos}
We next consider the effects of weight disorder alone, with no noise or delays ($g$ nonzero but $\sigma=d=0$ in \eqref{eq:predcode}).  This network has been shown to exhibit a dynamical phase transition from being a fixed point attractor when $g \leq g_c$ to chaotic evolution induced by large weight disorder for $g \geq g_c$ \cite{sompolinsky1988chaos}. The critical transition point $g_c$ depends on the nonlinearity $\phi$ and strength of inputs $x$. Roughly, higher nonlinear gains $\phi'(x)$ promote chaos by reducing $g_c$.  However, $g_c$ does not depend on the degree of balance where chaos and balance coexist \cite{Kadmon2015-rz,goedeke2017transition}. For $g \leq g_c$, there are no temporal fluctuations, so the only source of error is bias, which is computable and therefore can be removed.  Thus we focus on the chaotic regime $g \geq g_c$ in which the amplitude of chaotic fluctuations of membrane voltages $h_i(t)$ increases with $g-g_c$ \cite{Kadmon2015-rz}. In essence, the recurrent input  $g \eta_i(t) \equiv g \sum_j \mathcal J_{ij} \phi(h_i(t))$ due to the random connectivity $\mathcal J$ acts like a source of chaotic noise, analogous to the stochastic noise source $\sigma \xi_i(t)$ studied in Sec. \ref{sec:noise}.  A major difference however is that while the stochastic noise source is white across both neurons and time, with cross correlation $\langle \xi_i(t) \xi_j(t') \rangle = \delta_{ij} \delta(t-t')$,  the chaotic noise is, up to $O(1/\sqrt{N})$ corrections, white across neurons, but not across time, with cross correlation $\langle \eta_i(t) \eta_j(t') \rangle = \delta_{ij} q(t-t')$.  For chaotic models, the temporal autocorrelation function $q(t-t')$ must be solved self-consistently  \cite{Schuecker2018-gw,stapmanns2020self}, and within the chaotic regime it decays to a constant value on a time scale close to the membrane time constant $\tau$.  

While the full solution for the chaotic system is highly involved (see SM for comments on the derivation), we can describe the main quantitative effects of chaos on predictive coding error through an exceedingly simple derivation, which we give here.  Basically, we can account for the chaotic fluctuations simply by replacing the white noise $\sigma \xi_i(t)$ in \eqref{eq:predcode} by colored noise $g\eta_i(t)$ with temporal autocorrelation $q(t-t')= \exp(-\vert t-t'\vert/2\tau)$, which qualitatively matches the typical self-consistent solution to $q(t-t')$ in the chaotic regime. While this simplification does not describe the spatial structure of the chaos, which resides on a low-dimensional chaotic attractor \cite{Rajan2010nips}, it does capture the temporal structure of the chaos which, as we see next, primarily determines the error of balanced predictive coding. We then follow the noise based derivation in Sec. \ref{sec:noise}. The analog of \eqref{eq:phiprime} becomes 
\begin{equation}
\label{eq:chaoticfluct}
    \tau \delta \dot u(t) = -\delta u(t) - b \langle \phi' \rangle \delta u(t) + g \eta^{\parallel}(t) \quad \text{where} \quad \langle \eta^\parallel(t) \eta^\parallel(t') \rangle = \frac{1}{N} \exp(-\vert t-t'\vert/2\tau).  
\end{equation}
Thus the fluctuations $\delta u$ of membrane voltages $h_i(t)$ in the decoder direction $\mathbf w$ are well approximated by a leaky integrator with negative feedback proportional to $b \langle \phi' \rangle$ driven by colored noise, which is a stochastic ODE that is well understood \cite{hanggi1995colored}. 
Importantly, when the auto-correlation time of the driving noise equals the membrane time constant, as in this case, the variance is given by (see SM) $\langle \delta u^2\rangle\approx  \frac{g^2}{2N\langle\phi'\rangle^2b^2}$, yielding  a decoder variance 
\begin{equation*}
\langle \delta \hat x^2\rangle \approx \frac{\langle\phi'^2\rangle g^2}{2N\langle\phi'\rangle b^2N}.
\end{equation*} 
This should be compared to the decoder variance in \eqref{eq:secondorder} in the case of white noise, which instead scales as $O(\frac{\sigma^2}{bN})$. 
A rough intuition for the difference between chaos and noise can be obtained by considering the Fourier decomposition of the dynamics.  In the case of colored noise, the power of the fluctuations is concentrated at low frequencies, while for white noise it is evenly distributed across the spectrum and thus is spread thin. Increasing $b$ strengthens the filtering and suppresses more low-frequency fluctuations. As a result, when $b$ is increased by a fixed amount, the relative change in the power spectrum suppressed is higher, effectively improving the efficiency. The scaling of $\langle\delta u^2\rangle$ with the balance , as the exact inverse-quadratic power  $1/b^2$ is a result of the exact match between the time-constant, $\tau$  of the noise autocorrelation function $q(t-t')$ and of the dynamics in \eqref{eq:chaoticfluct} (see SM for details).
Thus our analysis reveals the important prediction that balance much more effectively suppresses decoder variance due to chaos versus noise, with a power law decay exponent in $b$ that {\it doubles} when going from noise to chaos.  We verify this important prediction in Fig. \ref{fig1}D.

\section{The role of delays, balance and noise in the onset of oscillatory instability}
\label{sec:delays}

In the previous two sections we have seen that increasing balance $b$ always suppresses decoder variance $\langle \delta \hat x^2\rangle$, for fluctuations induced both by noise and chaos. We now consider the case of a nonzero synaptic communication delay $d$, focusing first on the case of noise and no chaos (i.e. $d,\sigma >0$ and $g=0$ in \eqref{eq:predcode}). In this setting, the entire derivation of Sec. \ref{sec:noise} follows without modification until the analysis of membrane voltage fluctuation dynamics $\delta u(t)$ along the decoder direction $\mathbf w$ in \eqref{eq:phiprime}.  With a nonzero delay $d$, the dynamics of $\delta u(t)$ in \eqref{eq:phiprime} is modified to 
\begin{equation}
\label{eq:du_delay}
    \tau \delta \dot u(t) = -\delta u(t) - b \langle \phi' \rangle \delta u(t-d) + \sigma \xi^{\parallel}(t). 
\end{equation}
This corresponds to a delay differential equation \cite{gyHori1991oscillation}.  
We first consider its properties in the absence of noise input.  
First, for either zero balance $b$ or zero delay $d$, the dynamics has a stable fixed point at $\delta u = 0$. 
However, if either the delay $d$ is increased at fixed $b$, or the negative feedback $b$ is increased at fixed delay $d$,  the combination of strong negative feedback $b$ and long delay $d$ can trigger an oscillatory instability. 
To detect this instability, we search for complex exponential solutions to \eqref{eq:du_delay} of the form $\delta u(t) = e^{zt}$ where the complex frequency $z = \gamma + i\omega$.  
These solutions correspond to stable damped oscillations at frequency $\omega$ if $\gamma < 0$, or unstable diverging oscillations if $\gamma > 0$. 
Inserting $\delta u(t) = e^{zt}$ into \eqref{eq:du_delay} yields a constraint on $z$ through the characteristic equation $G(z) = z\tau+1+\tilde{b} e^{-zd}=0$ where $\tilde{b}\equiv b\langle\phi'\rangle$ is the \emph{effective} negative feedback taking into account the average nonlinear gain $\langle\phi'\rangle$ in \eqref{eq:phiprime}. 
At zero delay $d$, it has a solution $z = -(1+b)/\tau$ indicating damped exponential approach to the fixed point $\delta u = 0$. 

However, for a fixed delay $d$, as one increases the negative feedback $\tilde b$, the solutions $z$ to $G(z)=0$ move in the left half of the complex plane with negative real part $\gamma < 0$ towards the imaginary axis with $\gamma=0$.  
Let $\tilde b_c$ be the smallest, or critical value of $\tilde b$ for which $G(z)=0$ first acquires solutions on the imaginary axis, indicating the onset of oscillatory instability for any $\tilde b \geq \tilde b_c$. 
We can find $\tilde b_c$ by searching for solutions of the form $G(i \omega_c) = 0$. 
The real and imaginary parts of this complex equation yield two real equations: $\tilde{b}_c\cos(\omega_c d)+1=0$ and 
$\tilde b_c\sin(\omega_c d)-\omega_c\tau=0$. 
Here, $\omega_c$ is the frequency of unstable oscillations at onset, when $\tilde b$ approaches $\tilde b_c$ from below.  Solving for $\tilde b_c$ yields
\begin{equation}
\label{eq:bc}
\frac{d}{\tau}= {\arccos(-1/\tilde b_c)}/{\sqrt{\tilde b_c^2-1}}.
\end{equation}
Thus the maximal stable negative feedback $\tilde b_c$ is a function only of the relative delay $d/\tau$. Indeed $\tilde b_c$ is a decreasing function of $d$, indicating the longer the delay, the weaker the negative feedback must be to avoid oscillatory instabilities.  Beyond the linear oscillatory instability, with $\tilde b \geq \tilde b_c$, each neuron $i$ oscillates with amplitude proportional to $w_i$, stabilized by nonlinear saturation due to $\phi$.   

\begin{figure}
\centering
\includegraphics[width=\textwidth]{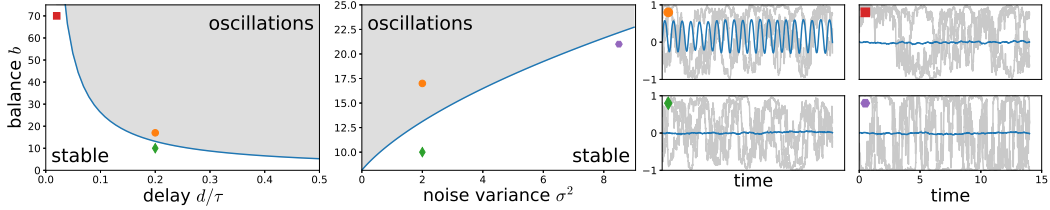} 
  \caption{\label{fig2_phase} Dynamical phases in the presence of delays, balance and noise ($g=0$, $x=0.2$). Left: The critical balance $b_c$ (blue curve) as a function of the delay (with $\sigma^2=2$) obtained by solving for $\tilde b_c$  in \eqref{eq:bc} and dividing by $\langle\phi'\rangle$ in \eqref{eq:phiprime}. Center:  The critical balance $b_c$  as a function of noise $\sigma$ for fixed delay ($d/\tau=0.15$). Right: sample firing rates $r_i(t)$ (grey) from simulations of \eqref{eq:predcode} with $N=1000$, with parameters corresponding to points in the left two panels, and the decoder trajectory $\hat x(t)$ (blue).}
\end{figure}

Importantly, the critical balance $b_c=\tilde b_c/\langle\phi'\rangle$ depends on the average gain of the nonlinearity $\langle\phi'\rangle$, which in turn depends on the degree of noise $\sigma$ through \eqref{eq:phiprime}.  Increasing $\sigma$ spreads out the distribution of membrane voltages $h_i(t)$ across neurons $i$.  For typical saturating nonlinearities, this {\it increased} spread in membrane voltages leads to a {\it decreased} average nonlinear gain, which in turn {\it raises} the critical balance level $b_c$, thereby allowing stronger negative feedback $b$ without triggering oscillatory instabilities.  Essentially, longer delays promote synchrony, while noise suppresses it, at any fixed balance. The predicted phase boundary between stable noise suppression and oscillatory amplification in the simultaneous presence of noise, delays and balance is verified in simulations (Fig. \ref{fig2_phase}).  
.

\section{An optimal level of balance in the face of noise, chaos and delays}
\label{sec:optimal}

\begin{figure}

\centering
\includegraphics[width=1.0\textwidth]{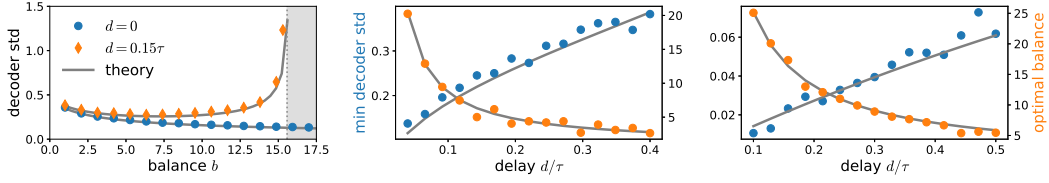} 
\caption{\label{fig3_optimal} Optimally balanced network with delay, $\phi=\tanh$ and $x=0$. Points reflect simulations of \eqref{eq:predcode} with $N=1400$ and curves reflect theory.
Left: Decoder standard deviation ($\sqrt{\langle (\delta \hat{x})^2\rangle}\times\sqrt{N}$) as a function of balance $b$ with $\sigma=0.75$. For $d=0$ this deviation decreases monotonically with $b$ as predicted by \eqref{eq:secondorder}. With nonzero $d$ this deviation exhibits a tradeoff between noise suppression and resonant amplification as predicted by \eqref{eq:fluc_w_delay}, with strong global oscillations triggered at $b\geq b_c$ (grey region), as predicted by \eqref{eq:bc}.  The optimal $b$ occurs at $b_{opt}=b_c/2$ (see text). 
Center: optimal decoder standard deviation (blue) and $b_{opt}$ (orange) as a function of delay, given by \eqref{eq:minerror} with $\sigma=0.75$. Asymptotically, the error increases as $\sqrt{d/\tau}$.
Right: Same as center but with deterministic chaos  ($g=1.6$, $\sigma=0$). Theory curves are calculated via colored noise (see Sec. \ref{sec:balancechaos}).} 
\end{figure}

We now examine how the presence of the oscillatory instability of the previous section impacts the nature of optimal predictive coding,  by considering how the delay dynamical system in \eqref{eq:du_delay} responds to the noise source $\sigma \xi^{\parallel}(t)$ in the stable regime, with $\tilde b \leq \tilde b_c$. We can understand the response in the frequency domain (see SM for detailed derivation). The power spectrum $\Delta(\omega)$ at frequency $\omega$ of the fluctuating time series $\delta u(t)$ can be written in terms of the characteristic function $G(z)$ as $\Delta(\omega)=[G(i\omega)G^*(i\omega)]^{-1}\sigma^2$, and the total variance is given by $\langle \delta u^2 \rangle = \int_{-\infty}^\infty d\omega \Delta(\omega)$. Now as $b$ approaches $b_c$ from below, the response power $\Delta(\omega_c)$ at the critical resonant frequency $\omega_c$ increases, since $G(i\omega_c)=0$ when $b=b_c$.  However, the power $\Delta(\omega)$ at non-resonant frequencies $\omega$ far from $\omega_c$ is suppressed by increasing $b$.  Indeed the total variance of both $\langle \delta u^2 \rangle$ and $\langle \delta \hat x^2 \rangle$ can be approximated by the sum of the power in the nonresonant frequencies, calculated above in \eqref{eq:secondorder}, and the power at the resonant frequency $\Delta(\omega_c)$, yielding (see SM)
\begin{equation}
\label{eq:fluc_w_delay}
\langle\delta \hat{x}^2\rangle=   \frac{\sigma^2\langle \phi'\rangle^2}{2\tau N}\left( \frac{1}{1+\tilde b}    
+ \frac{1}{\tilde b_c-\tilde b}\right).
\end{equation}
 This expression exhibits a tradeoff: increasing $b$ attenuates the first term by suppressing non-resonant input noise frequencies, but increases the second term by amplifying resonant noise frequencies.  Intriguingly, this fundamental tradeoff sets an optimal level of balance that minimizes decoder variance (Fig. \ref{fig3_optimal}).  Indeed minimizing \eqref{eq:fluc_w_delay} yields an optimal balance  $\tilde b_{opt}=\frac{1}{2}\tilde b_c$ (note $\langle\phi'\rangle$ does not depend on $b$ to leading order in $1/\sqrt{N}$). The resultant minimal error, $\varepsilon^2_{min}=\langle \delta \hat x^2 \rangle$ as a function of the delay is shown in Fig. \ref{fig3_optimal}.  For small delays $d\ll\tau$, the asymptotic expansion of \eqref{eq:bc} yields  $\tilde b_c\approx\pi\tau/2d$, and so the error increases initially as the square-root of the delay and is given by
\begin{equation}\label{eq:minerror}
\varepsilon_{min}  =  2\sigma\langle\phi'\rangle\sqrt{ \frac{ d}{N\tau\pi }}.
\end{equation}

\paragraph{Weight disorder, chaos and delays.}
 Delays do not change the statistics of chaotic fluctuations, since the mean-field equations are stationary, and fluctuations at times $t$ and $t-d$ are equivalent. Moreover, the maximal critical balance $\tilde b_c$ does not depend on the fluctuations and is still given by \eqref{eq:bc}. Below critically $b<b_c$ and for small delays $d\ll\tau$, resonant amplification at frequency $\omega_c$ plays less of a role in the case of chaos, since $\omega_c \propto 1/d$ and the power spectrum of chaotic fluctuations is exponentially suppressed at frequencies $\omega \gg 1/\tau$. Without a strong tradeoff between nonresonant suppression and resonant amplification, the optimal balance $b_{opt}$ for chaos is close to the maximal balance $b_c$, with a minimal decoder standard deviation that scales as $\varepsilon_{min}\propto 1/b_c$. For small delays where $b_c\sim\tau/d$, the minimal deviation scales as: $\varepsilon_{min}\sim d/\tau$. Our predicted scaling of optimal balance and deviation with delay in the case of chaos is confirmed in simulations (Fig. \ref{fig3_optimal}).

\section{Discussion}
In summary we have introduced a theoretically tractable nonlinear neural circuit model of predictive coding, and analytically derived many relations between coding accuracy and balance, noise, weight disorder, chaos, delays, and nonlinearities.  We find: (1) strong balance is a key requirement for superclassical error scaling with network size; (2) without delays, increasing balance always suppresses errors via powers laws with different exponents (-1 for noise, -2 for chaos); (3) delays yield an oscillatory instability and a tradeoff between noise suppression and resonant amplification; (4) this tradeoff sets a maximal critical balance level which decreases with delay; (5) noise or chaos can increase this maximal level by promoting desynchronization; (6) the competition between noise suppression and resonant amplification sets an optimal balance level that is half the maximal level in the case of noise; (7) but is close to the maximal level in the case of chaos for small delays, because the slow chaos has small power at the high resonant frequency; (8) the optimal decoder error rises as a power law with delay (with exponent 1/2 for noise and 1 for chaos).  Also, our model unifies a variety of perspectives in theoretical neuroscience, spanning classical synaptic balance \cite{Van_Vreeswijk1996-rm,Renart2007-kz,darshan2018strength,ebsch2018imbalanced,helias2014correlation,rosenbaum2014balanced}, efficient coding in tight balance \cite{Boerlin2013-ks, Deneve2017-lz},  the interplay of structured and random connectivity in computation \cite{Landau2018-jf,Mastrogiuseppe2018-sv,Mastrogiuseppe2018-hz,Boerlin2011-za,Ahmadian2013-xl}, the relation between oscillations and delays in  neural networks \cite{Lindner2005-oc,Roxin2005-hj,Brunel2006-xl} and predictive coding \cite{rao1999predictive, Deneve2016-jv}. Moreover, the mean-field theory developed here can be extended to spiking neurons with strong recurrent balance and delays \cite{Timcheck2020}, analytically explaining relations between delays, coding and oscillations observed in simulations but previously not understood \cite{Schwemmer2015-jp,Chalk2016-zr}   


\section*{Acknowledgments}
JK thanks the Swartz Foundation for Theoretical Neuroscience for funding; JT thanks the National Science Foundation for funding. SG thanks the Simons and James S. McDonnell Foundations and an NSF Career award for funding. We thank ID Landau and H Sompolinsky for the fruitful discussions.

\renewcommand{\appendixtocname}{Supplementary material}
\renewcommand{\appendixname}{Supplementary material}
  \renewcommand{\appendixpagename}{Supplementary material}
\begin{appendices}
\setcounter{equation}{0}
\renewcommand{\theequation}{S\arabic{equation}}

\section{Encoding an autonomous dynamical system}
\label{sec:stimuliMulti}


In the main text, we considered a case where a neural network encodes a scalar input signal $x(t)$ in its dynamics. This simple example corresponds to the circuit acting as an autoencoder. The model is instructive, and allows us to rigorously study the effects of  noise, weight disorder, and delays on the coding performance. In the simple case of an autoencoder, the desired output is explicitly provided to the network through the feedforward inputs. In a more general setting, the desired output of the network may be a complex spatiotemporal transformation of its input. The input-output transformation reflects the processing executed by the neural circuit, the details of which depends on the specific computation implemented. In this section, we show that the mechanism by which strong synaptic balance enables high-fidelity computations is general, and does not depend on origins of the signal. More precisely, we will show that the network can encode a linear dynamical system and that the resulting circuit equations obey similar balance rules as studied in the main text. Our assumption here is that the computed task can be written in terms of an autonomous linear dynamical system. Below we will also argue that this can be extended to other, nonlinear autonomous dynamical systems.

Our derivations follow closely those first suggested by \cite{Boerlin2013-ks} for spiking networks with integrate-and-fire dynamics. Here, we generalize the derivation to rate-based networks with arbitrary local nonlinear transfer functions. We explicitly show that the same ideas introduced in the theory of efficient coding in spiking neural networks \cite{Boerlin2013-ks,Deneve2016-jv} apply to continuous firing-rate models. Furthermore, we emphasize that the crucial component allowing the network to encode an arbitrary linear dynamical system is a decoder that introduces an additional time scale.

\subsection{Latent dynamical system}
Consider again a network of $N$ nonlinear neurons. The output of each neuron is given by a nonlinear transformation $\phi$ of its input. We wish to implement the arbitrary linear dynamics for the latent vector $\mathbf{x}(t)\in\mathbb{R}^M$,
\begin{equation}
\tau_x \dot{\mathbf{x}}(t) = \mathbf A \mathbf{x}(t) + \mathbf{x}_0(t).
\end{equation}
Here, $\mathbf{x}_0(t)\in\mathbb{R}^M$ is the input, e.g., from external stimuli, $\mathbf A$ is an arbitrary state transition matrix, and $\tau_x$ is the timescale. We assume that $N\gg M$; this is a fundamental assumption and it is needed in order to obtain the statistical benefits of distributed coding. We refer to $\mathbf x(t)$ as the \textit{latent} variable as it is not explicitly provided to the network, and its state is updated internally in the network.

The time scale of the dynamics $\tau_x$ can be very different than the microscopic timescale of the membrane potential, $\tau$.  In general, we expect the dynamical time scale of interest to be much longer than membrane potential $\tau_x\gg\tau$; the choice of slow dynamics is equivalent to the adiabatic limit used in the main text. We also note that inputs of arbitrary dimensionality $M'$ can be fed into dynamics of this form, simply by multiplying by an $M \times M'$ input matrix to correct the dimension.

As in the main text, a linear readout provides an estimator for the encoded variable $\hat{\mathbf{x}}(t)$. Unlike the autoencoder model, the estimate is of the latent variable $\mathbf{x}(t)$, and not of the direct input to the network $\mathbf{x}_0(t)$. Importantly, to realize a dynamical system, we need to introduce a timescale relevant for the encoded system $\tau_x$. This timescale can be introduced through the readout, 
\begin{equation}\label{eq:SM_r}
\tau_r\dot{\mathbf{r}}(t)=-\mathbf{r}(t)+\phi(\mathbf{h}(t)).
\end{equation}
Here $\mathbf{r}(t)$ is a smoothed version of the activity $\phi(\mathbf{h}(t))$ with a linear low-pass filter.  The slower dynamics of the readout provides the network with the necessary memory to implement the dynamics at slow time scales, even when the microscopic dynamics is fast and $\tau \ll \tau_x$. For brevity of our derivations, we let the readout time scale be $\tau_r=\tau_x$. In general, the readout and dynamics can have different time constants.  However, if the readout is too slow it will not capture the high frequencies in the dynamics. On the other hand, if the readout timescale is too fast it will not have the necessary memory to implement slow dynamics. Thus, the readout timescale needs to comparable to that of the latent dynamics. Any finite differences can be incorporated into the circuit equations. For an autoencoder, which has no latent dynamics, there is no need for introducing a slow timescale in the decoder. In this case we take the limit $\tau_r\to0$ leading to the simple relation $\mathbf{r}(t)=\phi(\mathbf h(t))$ used in the main text.

The estimator $\hat{\mathbf x}(t) $ of the latent state $\mathbf{x}(t)$ is obtained by a linear projection of the decoder rates $\mathbf r(t)$ onto the $M$-dimensional space of the latent dynamics
\begin{equation}
\hat{\mathbf{x}}(t) = W \mathbf{r}(t) = \frac{1}{N} \sum_{\alpha=1}^M \mathbf{w}_\alpha^T \mathbf{r}(t), \label{eq:readoutMulti}
\end{equation}
Here,  $\mathbf w_\alpha\in\mathbb{R}^N$  are the linear readout vectors, or coding directions. Each element in $\mathbf{w}_\alpha$ is drawn  i.i.d. from the same distribution $\mathcal{P}(w)$.  The coding directions are approximately orthogonal in the thermodynamic (large-$N$) limit, and the overlap between two different vectors is $\sum_iw_{\alpha i}w_{\beta_i}=O(1/\sqrt{N})$ for $\alpha\neq\beta$.  The distribution is normalized so that $\mathbf w_\alpha^T\mathbf w_\alpha=N$. It follows that  in the large $N$ limit $W^TW= \mathbf{I}$, where here $\mathbf{I}$ is the $M\times M$ identity matrix.  Together,  the readout vectors span the $M$ dimensional subspace of the latent dynamics. 

Following the paradigm of predictive coding \cite{rao1999predictive}, we want the internal state of the network to represent the error in estimation. The error vector, or deviation of the current estimate from the target latent state is $\mathbf x(t)-\hat{\mathbf x}(t)$. We thus define internal state variables, that we identify as the membrane potential of the neurons, which are equal to the projection of the error into the $N$-dimensional neural space

\begin{equation}\label{eq:SM_h}
\mathbf{h}(t) = b \left[ W^T (\mathbf{x}(t) - \hat{\mathbf{x}}(t) ) \right].
\end{equation}
Here, we have introduced an gain factor $b\in\mathbb{R}$, that defines the scale of the membrane potentials relative to the real readout error. We will soon identify this factor as the degree of balance in the network.  We would like the dynamics of the network to have a stable attractor around $\mathbf h=0$.  The outputs of the neurons are a nonlinear transformations of the membrane potentials, so the output of neuron $i$ is given by $\phi(h_i(t))$. 

In the analysis of the autoencoder in the main text,  the dynamical equations  of the membrane potentials $\mathbf{h}(t)$ were given \textit{a-priori} by a the canonical circuit equations \cite{amari1972characteristics}. Here, on the other hand,  the temporal evolution of $\mathbf{h}(t)$ is not independent, and is tied to the dynamics of the signal $\mathbf{x}(t)$ and of the readout $\hat{\mathbf{x}}(t)=W\mathbf{r}(t)$. To see how the membrane potentials evolve with time, we take a temporal derivative in both sides of \eqref{eq:SM_h}, which yields

\begin{align*}
\dot{\mathbf{h}}(t) &= b \left[ W^T (\dot{\mathbf{x}}(t) - \dot{\hat{\mathbf{x}}}(t) ) \right] \\ 
&= b \left[ W^T \left(\frac{1}{\tau_x}  A \mathbf{x}(t) + \frac{1}{\tau_x} \mathbf{x}_0(t) - W \dot{\mathbf{r}}(t) \right) \right] \\
&= b \left[ W^T \left(\frac{1}{\tau_x}  A W \mathbf{r}(t) + \frac{1}{\tau_x} \mathbf{x}_0(t) + \frac{1}{\tau_x} W \mathbf{r}(t) - \frac{1}{\tau_x} W \phi \left(\mathbf{h}(t) \right) \right) \right] & \label{eq:hdotfinalMulti}
\end{align*}
In the third step above we have used an approximation $\mathbf{x}\approx\hat{\mathbf x}(t)$; this approximation is valid as long as the readout error is small, and it  introduces an error of  $O(1/\sqrt{N})$ relative to the other $O(1)$ terms.  Rearranging the terms,  and absorbing the time constant $\tau_x$  within the free parameter $b$ we can write rewrite the dynamics as
\begin{equation}
\dot{\mathbf h}(t)= b \left[ W^T \mathbf{x}_0(t) -  W^T W \phi \left(\mathbf{h}(t) \right)
+  \Omega\mathbf{r}(t) 
\right].
\end{equation}
The first term is the input projected through feedforward weights. The second term is the inhibitory feedback implementing the error correction as we have seen in the main text. The last term is a recurrent term with weights given by $\Omega \equiv W^T ( A + I) W$, where $I$ is the $N\times M$ identity matrix. Importantly, this feedback term is proportional to the decoder rates $\mathbf{r}(t)$, and not the output of the neurons $\phi(\mathbf h)$. This is the term that implements the dynamics of $\mathbf x(t)$. It can be readily understood as it is the only term that contains the dynamic transfer matrix $A$ and the additional time constant, which is implicit inside the filtered readout $\mathbf r(t)$. In \cite{Boerlin2013-ks} they refer to these synapses as \textit{slow} synapses, as they inherit the slow dynamics of the readout $\mathbf r(t)$. In general, this term is a temporal filter of the neural outputs, that introduces a longer time scale required to implement the encoded dynamical system. 

Finally, to arrive at the full circuit equations analogous to \eqref{eq:circuit} in the main text, we introduce membrane leak, weight disorder, added Gaussian noise, and delays:  

\begin{equation}
\tau\dot{h}_i(t)=-h_i(t)+\sum_jJ_{ij} \phi(h_j(t-d)) + I_i(\mathbf{x}_0(t)) + b\Omega\mathbf{r}(t)+\sigma\xi_i(t), \label{eq:circuitMulti}
\end{equation}where
\begin{equation}
\label{eq:connectivityMulti}
    J_{ij} =  g \mathcal J_{ij}  -  \frac{b}{N} \sum_{\alpha=1}^M w_{\alpha i} w_{\alpha j},  \qquad \text{and} \qquad   I_i(\mathbf{x}_0(t)) = b \sum_{\alpha=1}^M w_{\alpha i} x_{0 \alpha}(t).
\end{equation}
Once again, we have absorbed a factor of $N$ within the arbitrary control factor $b$. The delays, noise, weight disorder, and leak are not part of the derivation, but can be seen as external constraints on the network. With the addition of the noise and disorder, we can naturally see the role of balance in the dynamics. It sets the effective scale of the error relative to the other driving forces in the network, which are the noise $\sigma$, and the emergent fluctuations due to the disorder, which are proportional to $g$. The mean-field derivation in the main text shows how the magnitude of $b$ affects the different sources of fluctuations in the network. 

A noticeable difference from the mean-field equations for the autoencoder, is the added term $b\Omega \mathbf r(t)$,that in general changes the result of the mean-field derivation. However, in the limit where the latent dynamics is much slower than the membrane time constant, and both $\tau_x,\,\tau_r\ll\tau$, then the fluctuations in the decoder rates  $\delta\mathbf{r}(t)=\mathbf{r}(t)-\langle\mathbf{r}\rangle$ are small and do not contribute to the overall fluctuations $\delta u(t)$  in \eqref{eq:secondorder}. On the other hand,  the contribution of the mean rates $b\Omega\langle\mathbf{r}\rangle$ will affect the bias in general.

We note that the linear dynamics can be generalized to nonlinear dynamics, by explicitly introducing nonlinearity within the readout \eqref{eq:SM_r}, and adapting the recurrent weights $\Omega$ accordingly. Similar ideas have been previously introduced in  \cite{Alemi2018-cj}.

Finally, to get the autoencoder network studied in the main text, we can choose $A = -I$, yielding $\mathbf{x}(t) = \mathbf{x}_0$. With this choice of $A$, $\Omega = 0$, and the ``slow'' recurrent connectivity term in Equation \eqref{eq:circuitMulti} drops out, leading to the circuit  equations introduced in \eqref{eq:circuit}, only for $M$-dimensional signals instead of a scalar input. In the following section, we derive the full mean-field theory for an autoencoder for an input signal of $M$ dimensions, where  $1<M\ll N$.

\section{Mean field theory for multidimensional stimuli}

In section \ref{sec:ErrorScale} of  the main text, we calculate the variance of fluctuations in a scalar readout; here we generalize the calculation of variance to multidimensional stimuli  by continuing from Equations \eqref{eq:circuitMulti} and \eqref{eq:connectivityMulti}. We will consider the more simple case of a network with no weight disorder and no delay,  $g = d = 0$. Futhremore, as in the main text we consider an autoencoder without internal signal dynamics, i.e., $A=-I$ and $\Omega=0$.

We define the readout in the direction $\alpha=1,\dots,M$ as
\begin{equation}
\hat{x}_\alpha(t) = \frac{1}{N} \mathbf{w}_\alpha^T \mathbf{r}(t).
\end{equation}
The dynamical fluctuations in the readout are given by 
\begin{equation}
\delta\hat{x}_\alpha(t)=\hat{x}_\alpha(t)-\langle\hat{x}_\alpha\rangle.
\end{equation}
The readout error is determined by the bias and the variance of the readout. Below we show that the fluctuations in the readout $\delta\hat{x}(t)$  in each direction $\alpha=1,\ldots,M$ are independent and so the total error can be written as
\begin{equation}\label{eq:MultiDerror}
\varepsilon = \sqrt{\sum_\alpha^M (x_\alpha -\langle\hat{x}_\alpha\rangle)^2+\sum_\alpha^M\langle\delta\hat{x}_\alpha^2\rangle},
\end{equation}
where the first term in the square root is the contribution of the bias, and the second term is the contribution of the variance of dynamical fluctuations.

\subsection{First-order mean field theory for the bias}
Following the same mean-field analysis as in the main text, we decompose the membrane voltage vector $\mathbf{h}$ into two contributions 
\begin{equation}
\mathbf h(t) = \sum_{\alpha=1}^M \mathbf h_\alpha^{\parallel}(t) + \mathbf h^{\perp}(t).
\end{equation} 
Here  $\mathbf h_\alpha^{\parallel} = \mathbfcal P_\alpha \mathbf h$  and $\mathbf h^{\perp} = (\mathbf I - \sum_{\alpha=1}^M \mathbfcal P_\alpha) \mathbf h$ are the projections of the membrane potential vector onto the subspace spanned by $\{\mathbf w_\alpha\}$ and to the orthogonal subspace respectively.  $\mathbfcal P_\alpha = \frac{1}{N} \mathbf w_\alpha \mathbf w_\alpha^T$ is the orthogonal projection operator. Importantly, the readout vectors $\mathbf{w}_\alpha$ are approximately orthogonal in the large $N$ limit, enabling this decomposition. Thus $\mathbf h_\alpha^{\parallel}(t) = u_\alpha(t) \mathbf w_\alpha$ where $u_\alpha(t) \equiv \frac{1}{N} \sum_{i=1}^N w_{\alpha i} h_i(t)$.  The two dynamical equations \eqref{eq:decompose} in the main paper generalize to $M$ equations for the projections of the membrane potentials onto the subspace spanned by the readout vectors 
\begin{equation}
\label{eq:decomposeMulti}
\tau \dot u_\alpha(t) = -u_\alpha(t) + b \left[ x_\alpha - \hat x_\alpha(t) \right] + \sigma \xi_\alpha^{\parallel}(t),
\end{equation}
and for the fluctuations in the orthogonal subspace
\begin{equation}
\tau \dot h^\perp_i(t)  =  - h^\perp_i(t) + \sigma \xi^{\perp}_i(t).
\end{equation}
The noise terms  $\xi_\alpha^\parallel = \frac{1}{N} \sum_{i=1}^N w_{\alpha i} \xi_i$ reflect the projection of the single-neuron independent noise terms into the $\alpha$ readout direction, and $\xi^\perp_i$ is the independent noise in neuron $i$ in the orthogonal subspace. Since $M\ll N$, we can write, as in the main text,    $\langle(\xi_i^\perp)^2\rangle=\sigma^2$ and $(\xi^\parallel_\alpha)^2=\frac{\sigma^2}{N}$.  Additionally, since the coding directions  $\mathbf{w}_\alpha$ are approximately orthogonal, the $\xi_\alpha^\parallel$ are independent and $\langle\xi^\parallel_\alpha\xi^\parallel_{\beta}\rangle= 0$ for every pair $\alpha\neq\beta$.

The membrane potentials in the subspace orthogonal to all the coding directions, $\mathbf h^\perp(t)$   follow a simple OU process, and the variance of their fluctuations is given by $\langle (h^\perp_i)^2 \rangle = \frac{\sigma^2}{2\tau}$. Since the fluctuations in \textit{all} readout directions $\delta u_\alpha(t)$ are small in the large $N$ limit, we can expand the activity of each neuron to linear order in these fluctuations
\begin{equation}
\phi(h_i(t))= \phi \left( \sum_{\beta=1}^M w_{\beta i} \langle u_\beta \rangle + h^\perp_i(t)\right)+
\phi' \left( \sum_{\beta=1}^M w_{\beta i} \langle u_\beta \rangle + h^\perp_i(t)\right)\sum_{\beta=1}^M w_{\beta i}\delta u_\beta(t).
\end{equation}

The decoder $\hat x_\alpha(t)=N^{-1}\sum_i w_{\alpha i}\phi(h_i(t))$ then reads 
\begin{multline}
\label{eq:linearhatxMulti}
    \hat x_\alpha(t) = \frac{1}{N} \sum_{i=1}^N w_{\alpha i} \phi \left( \sum_{\beta=1}^M w_{\beta i} \langle u_\beta \rangle + h^\perp_i(t)\right) \\ + 
    \frac{1}{N} \sum_{i=1}^N w_{\alpha i}  \phi' \left( \sum_{\beta=1}^M w_{\beta i} \langle u_\beta \rangle + h^\perp_i(t)\right)\sum_{\beta=1}^M w_{\beta i}\delta u_\beta(t).
\end{multline}

Mirroring the derivation in the main paper, we take a temporal average of the decoder, and use $\langle\hat x_\alpha\rangle=x_\alpha-\langle u_\alpha\rangle/b$  , to obtain a set of $M$ self-consistent equations for the order parameters $\langle u_\alpha\rangle$,
\begin{equation}
\label{eq:firstorderMulti}
    x_\alpha - \frac{\langle u_\alpha \rangle}{b} =
    \int \mathcal Dz \, \prod_{\beta=1}^M (dw_\beta \mathcal{P}( w_\beta )) \, w_\alpha
  \phi \left( 
    \sum_{\beta=1}^M w_\beta \langle u_\beta \rangle  
    + \frac{\sigma}{\sqrt{2\tau}} z \right).
\end{equation}
These equations can be solved numerically to give the stationary solutions for $\langle u_\alpha\rangle$. In the main text we highlight an intuitive graphical solution. While the basic idea is similar in the multidimensional setting, the graphical solution is less intuitive since the  LHS of \eqref{eq:firstorderMulti} is a nonlinear integral equation involving  all of the order parameters $\langle u_\alpha\rangle$. 

In the mean-field solution the bias is the Euclidean distance between $\langle\hat{\mathbf x}\rangle $  and $\mathbf x$, given by
\begin{equation}
\varepsilon_{bias}=\frac{1}{b}\sqrt{\sum_\alpha^M\langle u_\alpha\rangle^2}.
\end{equation}
The bias $u_\alpha(\mathbf x,\sigma)/b$  is a function of the noise and the inputs $x_\alpha$ in all directions $\alpha=1,\ldots,M$. This is a deterministic function that can be inverted by, for example, training of an efferent readout which  can  eliminate the error due to bias. In the next section, we calculate the error due to dynamical fluctuations in the different coding directions. These depend on the noise and chaos in the network and are not easily removed by a static readout.

\subsection{Mean-field theory for the second order statistics of the fluctuations}
We now turn to study the fluctuations around the static first-order mean-field solution. By removing the time average from the expansion in \eqref{eq:linearhatxMulti}, we identify the fluctuations in the readout as
\begin{equation}
\delta \hat{x}_\alpha(t) =     \frac{1}{N} \sum_{i=1}^N w_{\alpha i}  \phi' \left( \sum_{\beta=1}^M w_{\beta i} \langle u_\beta \rangle + h^\perp_i(t)\right)\sum_{\beta=1}^M w_{\beta i}\delta u_\beta(t).
\end{equation}
As we have noted above, the coding directions are all random and  $\frac{1}{N}\sum_i w_{\alpha i}w_{\beta i}=\delta_{\alpha\beta}$. As a result the fluctuations $\delta u_\alpha(t)$  in different directions decouple and  follow the linear dynamics
\begin{equation}
    \tau \delta \dot u_\alpha (t) = -\delta u_\alpha (t) - b \langle \phi' \rangle_\alpha \delta u_\alpha (t) + \sigma \xi_\alpha^{\parallel}(t).
\end{equation}
Here, the average  $\langle\phi'\rangle$  is performed the statistics of the stationary solution calculated above, and depends in the means $\langle u_\alpha\rangle$ in all $M$ directions,
\begin{equation}\label{eq:phiprimeMulti}
    \langle \phi' \rangle_\alpha =
     \int \mathcal Dz \,\prod_{\beta=1}^M (dw_\beta  \, 
    \mathcal P(w_\beta) ) w_\alpha^2 
    \phi' \left(
    \sum_{\beta = 1}^M w_\beta \langle u_\beta \rangle
    + \frac{\sigma}{\sqrt{2\tau}} z \right).
\end{equation}

Since fluctuations decouple, we can solve the equation in each direction $\alpha$ independently, and the fluctuations in each direction are given by  
\begin{equation}
\label{eq:secondorderMulti}
    \langle (\delta u_\alpha(t))^2 \rangle  = \frac{\sigma^2}{2\tau N(1+b \langle \phi' \rangle_\alpha)},  \qquad \text{and}  \qquad
 \langle (\delta \hat x_\alpha (t))^2 \rangle = \frac{\langle \phi' \rangle_\alpha^2 \sigma^2}{2\tau N(1+b \langle \phi' \rangle_\alpha)}.
 \end{equation}
Finally, since the fluctuations are orthogonal and independent, the total  contribution of the fluctuations to the readout error is given by $\sqrt{\Delta}$, where $\Delta$ is variance of the decoder across all readout directions 
\begin{equation}
{\Delta}={\sum_\alpha^M\langle(\delta\hat x_\alpha(t))^2\rangle}.
\end{equation}
  %

\section{Dynamic mean-field theory for balanced networks with weight disorder}
In section \ref{sec:balancechaos} of the main text we study the effect of weight disorder and deterministic chaos on the error, and show how balance suppress the fluctuations at the readout. Here, we present with more details the mean-field solutions for chaotic networks with synaptic balance, and the approximations we introduced in order to study the effects of the balance on the dynamics. For simplicity, we derive the solutions here assuming a scalar input signal, as introduced in the main text. Furthermore, for notational brevity we set the membrane time constant to be $\tau=1$.

First, we note that the first-order mean-field solution for the bias \eqref{eq:firstorder} is unaffected by the dynamics of the noise, and is similar whether the fluctuations of the membrane potential arise from deterministic chaos or from additive Gaussian noise. However, the mean-field solution requires averaging over the membrane potential fluctuations in the directions orthogonal to the readout, which in general may be different in the case of deterministic chaos.  In the case of additive Gaussian noise, we have shown that the  temporal average of the fluctuations are  $\langle\delta (h^\perp_i)^2\rangle=\sigma^2/2$ for  all $i$. When the fluctuations are the result of deterministic chaos, the variance $\langle (h_i^\perp)^2\rangle$ is found self consistently via  Dynamic Mean Field Theory (DMFT) \cite{sompolinsky1988chaos}. In the following section, we highlight the main ideas in deriving DMFT for the emergent fluctuations in the membrane potential.

\subsection{Dynamic mean-field solution for the fluctuations in the orthogonal subspace}
We now turn to compute the statistics of the fluctuations of a random network in its chaotic phase, when the variance of the weight distribution is above the critical transition point $g>g_c$. The dynamic mean field theory for a chaotic neural network was first introduced by \cite{sompolinsky1988chaos} and re-derived later by \cite{Kadmon2015-rz,stapmanns2020self,goedeke2017transition,helias2019statistical,Mastrogiuseppe2018-hz,crisanti2018path}.  The connectivity in the subspace orthogonal to the readout direction is randomly distributed, thus the properties of the fluctuations in this subspace, $\delta h^\perp(t)$, are equivalent to previous studies of random neural networks. We bring the highlights  here, and refer the reader to \cite{Kadmon2015-rz} for a more detailed account of the derivation.

We define the autocorrelation function of the  chaotic fluctuations as
\begin{equation}\label{eq:DeltaPerp}
  \Delta^\perp(s)\equiv \frac{1}{N}\sum_i\left\langle \delta h^\perp_i(t) \, \delta h^\perp_i(t+s)\right\rangle
\end{equation}

 where, as before,  $\delta\mathbf{h}^\perp(t)=(I-\mathbfcal P)\delta\mathbf{h}(t)$.    The variance of the fluctuations is given by the equal-time autocorrelation $\Delta^\perp(0)$. In  DMFT, the autocorrelation is obtained by properly averaging over the dynamic equation for the fluctuations $\delta h_i^\perp(t)$, given by the equation in the RHS of\eqref{eq:decompose}. The result, is a second-order differential equation for $\Delta^\perp(s)$ given by
 
\begin{equation}\label{eq:ChaosDMFT}
  \left( 1-\frac{\partial^2}{\partial s^2} \right)\Delta^\perp(s)=g^2q(s).
\end{equation}

Here on the LHS we have a second-order differential operator acting on the autocorrelations of the membrane potential. On the RHS, we have the autocorrelation function of the fluctuations in the firing rates of the neurons  $\phi_i(t)\equiv\phi(h_i(t))$,  is given by 
\begin{equation}
  q(s)=\frac{1}{N}\sum_i\left\langle \delta\phi_i(t)\delta\phi_i(t+s)  \right\rangle.
\end{equation}
Here  $\delta\phi_i(t)=\phi_i(t)-\langle\phi_i\rangle$ are the temporal fluctuations in the output of  neuron $i$ about its mean firing rate $\langle\phi_i\rangle$.  The mean autocorrelation of the firing rates $q(s)$  is given by taking a statistical average over the weight disorder in the system, and can be written as

\begin{equation}\label{eq:qs}
  q(s)=\int Dz \left( \int Dy\phi(\sqrt{\Delta^\perp(0)-\Delta^{\perp}(s)}y+\sqrt{\Delta^\perp(s)}z\right)^2.
\end{equation}

Plugging \eqref{eq:qs} into \eqref{eq:ChaosDMFT} we get a self-consistent integro-differential equation for $\Delta^{\perp}(t)$. The boundary conditions for this equation are given by   $\dot{\Delta}^\perp(s)=0$ and $\dot{\Delta}^\perp(\infty)=0$, corresponding to the smoothness of the autocorrelation at $s=0$ and the conditions for the existence of a chaotic solution at $s=\infty$ respectively.   The solution can be found by numerically evaluating the second order differential equation \cite{Kadmon2015-rz}. The variance of the fluctuations in the orthogonal subspace $N^{-1}\sum_i\langle\delta h_i^{\perp2}\rangle=\Delta^\perp(0)$  is used  in the static solutions $\langle\phi\rangle$ and $\langle\phi'\rangle$ above.

\subsection{Dynamic mean-field for the fluctuations in the readout direction}
The dynamics of the fluctuations in the direction of the readout is given by
\begin{equation}\label{eq:du}
\tau \delta \dot u(t) = -\beta\delta u(t)  + g\eta^{\parallel}(t),
\end{equation} 
where $\beta\equiv1+b\langle\phi'\rangle$. Here, the noise term $\eta^\parallel(t)$ reflects the projection of the  recurrent feedback $\mathbfcal{J}\phi$ with random connectivity $\mathbfcal J$ onto the coding direction $\mathbf{w}$. The mean of the recurrent noise $\eta^\parallel(t)$ is given by  

\begin{equation}
\langle \eta^\parallel(t)\rangle = \frac{1}{ N}\sum_{ij} w_i\mathcal J_{ij}\langle \phi(h_i(t)) \rangle =\frac{a_{\mathcal J}}{\sqrt{N}}\langle\phi\rangle,
\end{equation}
where $a_{\mathcal{J}}\sim\mathcal{N}(0,1)$  is a random number drawn from the standard normal distribution. The random number depends on the particular realization of $\mathbfcal{J}$ and readout vector $\mathbf{w}$, and does not vanish in the large $N$ limit. Requiring \textit{detailed-balance} in the disordered connectivity, i.e., the constraint $\sum_j\mathcal{J}_{ij}=0,\;\forall i$ can remove this bias term. Without detailed-balanced weights, the realization-specific temporal mean needs be incorporated within the mean-field equation for  $u$, and will generally add to the bias error.  We note that the expected  $a_\mathcal{J}$ across different readout directions is zero.  As we argued before, the static bias can be removed by an efferent readout. However, in the case of weight disorder, the bias term is random and depends on the actual realization of $\mathbfcal{J}$, and there is no analytical solution for the bias. Nevertheless, the static bias can be easily removed by training the linear readout. The bias correction to the mean-field is needed even for the dynamical phase below the chaotic transition, $g<g_c$.

For networks in the chaotic phase, we must also consider the temporal fluctuations. The  autocorrelation of the noise term in \eqref{eq:du} is given by

\begin{equation}\label{eq:ChaosNoiseVar}
\langle \delta\eta^\parallel(t)\delta\eta^\parallel(t')\rangle =
\frac{1}{N^2}\sum_{ijkl}w_iw_j\mathcal J_{ik}\mathcal J_{jl}\langle\delta\phi_k(t)\delta\phi_l(t')\rangle
=\frac{1}{N} q(t-t')+O(1/N^2)
\end{equation}
where  $q(t-t')$ is the mean autocorrelation of the outputs given in \eqref{eq:qs}, which can be found self-consistently as highlighted above. Unlike the mean, it is self-averaging, and does not depend on the specific realization of $\mathbfcal{J}$ in the large-$N$ limit.

To  find an expression for the autocorrelation function 
\begin{equation}
\Delta(s)=\langle\delta u(t)\delta u(t+s)\rangle,
\end{equation} 
we follow the same logic as when deriving Dynamic Mean-Field Theory for the fluctuations $\Delta^\perp(s)$ above \cite{Kadmon2015-rz}.  First, we take the Fourier transform of the dynamical equations  \eqref{eq:du} for the fluctuations in the readout direction $\delta u(t)$:
\begin{equation}
(-i\omega-1-\beta)\delta\tilde{u}(\omega)={g}\eta^\parallel(\omega).
\end{equation}
Next, we multiply the expression by its complex conjugate and take another Fourier transformation back to the temporal representation.  Replacing the variance of $\eta^\parallel\rangle$ with the variance of the recurrent connectivity \eqref{eq:ChaosNoiseVar} we obtain

\begin{equation}\label{eq:DeltaDMFT}
  \left( (1+b\langle\phi'\rangle)^2-\frac{\partial^2}{\partial s^2} \right)\Delta(s)={g^2}q(s).
\end{equation}
The boundary conditions on the second-order differential equation are, as above in  equation \eqref{eq:ChaosDMFT}, are $\dot{\Delta}(0)=\dot{\Delta}(\infty)=0$. The full solution for $\Delta(s)$ can be evaluated numerically using the solution for $q(s)$ describes in the previous section. However, to get further insight into how the solution behaves with $b$, we would like to derive an analytical expression. In the following section, we approximate the chaotic autocorrelation function with a more simple model with a colored Gaussian noise term that permits analytical treatment.

\subsection{Approximating the chaotic fluctuations with temporally colored Gaussian noise}
The exact temporal correlation function of the chaotic fluctuations is complicated, and depends on the details of the problem, such as the nonlinearity, sources of noise and the external input \cite{Kadmon2015-rz,stapmanns2020self}. However, it has some common characteristics: (1) it is a symmetric function $q(s)=q(-s)$; this is due to time reversal symmetry in the system. (2) It is an exponentially decaying function; this is because the chaotic dynamics is characterized by a positive Lyapunov exponent \cite{engelken2020lyapunov}. (3) The decay time is of the order of the membrane potential, which is the only time scale in the network. The last point is true away from the critical transition point $g=g_c$, where critical slowing down can result in long-range temporal correlations \cite{Kadmon2015-rz}.  The exact shape of the autocorrelation however, depends on the details of the problem. For example it may be convex or concave, depending on external noise sources \cite{stapmanns2020self}.

While the detailed function is not analytically tractable in many cases, we can replace the chaotic fluctuations with a more simple noise model that captures the important aspects of the chaotic fluctuations, namely,  symmetric and exponentially decaying with time constant similar to the membrane time constant. We write the dynamics of the fluctuations in the direction of the readout in \eqref{eq:du} as 
\begin{equation}\label{eq:ChaosZeta}
\tau \dot{\delta u}(t)=-\beta\delta u(t) + g\zeta(t),
\end{equation}
where $\beta=1+b\langle\phi'\rangle$. Here, we have replaced the chaotic fluctuations in the coding direction, $\eta^g(t)$ with correlated Gaussian noise $\zeta(t)$ with zero mean and autocorrelation function given by 
\begin{equation}
  \langle\zeta(t)\zeta(t+s)\rangle=\frac{1}{N}\exp\left(-\frac{\vert s\vert}{2\tau}\right).
\end{equation}
This noise can be easily realized with a filtered white Gaussian noise 
\begin{equation}
  \tau \dot{\zeta}(t)=-\zeta(t)+\frac{1}{\sqrt{N}}\zeta'(t),
\end{equation}
where $\langle\zeta(t)\zeta(t')\rangle=\delta(t-t')$. For brevity of notation, in the following we will set membrane time constant $\tau=1$.

 In \eqref{eq:ChaosZeta} we have  a stochastic ODE with a corresponding to a particle undergoing gradient descent in a deterministic quadratic potential, but driven by colored noise. If at time $t_{0}$ the location of the particle
is known, then the variance in the location of the particle  at time $t$ is given by \cite{gyHori1991oscillation}
\begin{equation}
\alpha(t_{0},t)=2\int_{t}^{t'}ds\exp[-2\beta(t_{0}-s)]\int_{t_{0}}^{s}drC(s-r)\exp[-\beta(s-r)].
\end{equation}
Here $C(s-r)$ is the autocorrelation function of the colored Gaussian driving noise, and is given by $C(s-r)=\frac{1}{N}\exp[-(s-r)/2]$ for $t'>t$. We thus obtain 
\begin{multline}
\alpha(t_{0},t)=\frac{2}{N}\int_{t_{0}}^{t}ds\exp[-2\beta(t-s)]\int_{t_{0}}^{s}dr\exp[-(\beta+\frac{1}{2})(s-r)]\\
=\frac{2}{N(\beta+\frac{1}{2})}\int_{t_{0}}^{t}ds\exp[-2\beta(t-s)]\left(1-\exp[-(\beta+\frac{1}{2})(s-t_{0})]\right)\\
=\frac{\exp[-2\beta t]}{N(\beta+\frac{1}{2})}\int_{t_{0}}^{t}ds\left(\exp[2\beta s)]-\exp[(\beta+\frac{1}{2})t_{0}]\exp[(\beta-\frac{1}{2})s]\right).
\end{multline}

If the balance is strong, we can write $\beta=b\langle\phi'\rangle\gg1$. This small approximation allows us to simplify the above expression by ignoring $O(1)$ corrections to $\beta$, and write  
\begin{multline}
\alpha(t_0,t)=\frac{\exp[-2\beta t]}{N\beta}\int_{t_{0}}^{t}ds\left(\exp[2\beta s)]-\exp[\beta t_{0}]\exp[\beta s]\right)\\
=\frac{\exp[-2\beta t]}{N\beta^{2}}\left(\frac{1}{2}\left(e^{2\beta t}-e^{2\beta t_{0}}\right)-e^{\beta t_{0}}\left(e^{\beta t}-e^{\beta t_{0}}\right)\right)\\
=\frac{1}{N\beta^{2}}\left(\frac{1}{2}\left(1-e^{-2\beta(t-t_{0})}\right)-e^{\beta t_{0}}\left(e^{-\beta t}-e^{\beta t_{0}-2\beta t}\right)\right)\\
=\frac{1}{2N\beta^{2}}\left(\left(1-e^{-2\beta(t-t_{0})}\right)-2\left(e^{-\beta(t-t_{0})}-e^{-2\beta(t-t_{0})}\right)\right).
\end{multline}

As mentioned above, if we interpret the ODE as the motion of a particle in a quadratic potential driven by colored noise, then  $\alpha(t_0,t)$ denotes the variance in the location of the particle at time $t$, if the location is known at time $t_0$. In that case the variance in the location of the particle at the \textit{steady state} is given by setting $t_0=0$ and $t=\infty$, yielding
\begin{equation}
\alpha(0,\infty)=\frac{1}{N2\beta^{2}}=\frac{1}{2N\tilde{b}^{2}}=\frac{1}{2N\langle\phi'\rangle^{2}b^{2}}.
\end{equation}
Here we have again used $\tilde{b}\gg1$ so  $\beta=1+\langle\phi'\rangle b\approx\langle\phi'\rangle b$.

Using the result for the variance at the steady state, we obtain an expression for the fluctuations in $\delta u(t)$ 
\begin{equation}
\langle \delta u^2\rangle\approx  \frac{g^2}{ 2\tilde{b}^2N}.
\end{equation}

Finally, the fluctuations of the readout $\hat{x}(t)$ are given by
\begin{equation}
\langle  \delta \hat{x}^2\rangle\approx  \frac{\langle \phi'\rangle^2  g^2}{ 2\tilde{b}^2N}.
\end{equation}
The average over the steady state $\langle \phi'\rangle$ is solved using the mean-field equations, using the variance $\Delta^\perp(0)$ found above using dynamic mean-field theory.

\section{Delays, noise and resonance}
In this section, we study the response of a balanced network with delayed feedback to an external white noise. We begin by considering the characteristic equation of the delayed ODE in \eqref{eq:du_delay} in the absence of noise, 
\begin{equation}
G(z)=z\tau+1+\tilde{b}e^{-zD}=0.
\end{equation}
The real and imaginary parts of the complex number  $z=\gamma+i\omega$ represent the exponential growth and oscillations of the solution ansatz. As discussed in the main text, below the critical balance $\tilde{b}_c$, all solutions to this equation have negative real part $\gamma<0$.  In this regime the dynamics is stable and the fluctuations decay to zero rapidly. 

In the presence of noise, the system is constantly driven. The autocorrelation function of the fluctuations in this state is defined as $\Delta(s)=\langle\delta u(t)\,\delta u(t+s)\rangle$. To study the response of of $\delta u(t)$ to the external noise, we look at the Fourier components of the autocorrelation function
\begin{equation}
\hat\Delta(\omega)=\frac{1}{2\pi}\int ds\,e^{i\omega s}\Delta(s).
\end{equation}
In the model driven by white noise, the integrated power across all frequencies is $\sigma^2/2N$. The power at a specific frequency $\omega$ is given by

\begin{equation}\label{eq:DeltaOmega}
\hat{\Delta}(\omega)=\frac{\sigma^2}{2NG(i\omega)G^*(i\omega)}=\frac{\sigma^2}{2N(i\omega\tau+1+\tilde{b}e^{-i\omega D})(-i\omega\tau+1+\tilde{b}e^{i\omega D})}.
\end{equation}
Using the characteristic equation we know that $G(\omega_c)$=0 when the balance is $\tilde b=\tilde b_c$. Plugging this equality into \eqref{eq:DeltaOmega}, we obtain an expression for the response of the network at the  resonant frequency $\omega_c$, 
\begin{equation}
\hat\Delta(\omega_c)=\frac{\sigma^2}{2N(\tilde{b}_c-\tilde{b})^2}.
\end{equation}
We approximate the total contribution to the fluctuations as the sum of fluctuations in the absence of delay plus the contribution of the resonance in $\omega_c$, yielding
 \begin{equation}\label{eq:tildeDelta}
 \hat\Delta(\omega)=\frac{\sigma^2}{2N}\left(
 \frac{1}{(1-\tilde{b}_c)^2+\omega^2}+\frac{1}{(\tilde{b}_c-\tilde{b})^2+(\omega-\omega_c)^2}
 \right).
 \end{equation}
 Finally, using the Wiener–Khinchin theorem we can find the total variance of the fluctuations, which is given by
 \begin{equation}\label{eq:WKT}
 \Delta=\int d\omega \hat{\Delta}(\omega).
 \end{equation}
 Plugging \eqref{eq:tildeDelta} in \eqref{eq:WKT} and integrating, we arrive at eq. \eqref{eq:fluc_w_delay}.
 
 Finally, we note that for the chaotic network, the derivation would be similar, only the variance of the white noise $\sigma^2/N$ is replaced with $\hat{q}(\omega)/N$, which is the Fourier representation of the rate autocorrelation in \eqref{eq:qs} . Since $q(s)$ is exponentially decaying, we have $\hat{q}(\omega)\ll1$ for $\omega\gg1/\tau$. In the case of small delays $d\ll\tau$ the the noise at the critical frequency $\hat{q}(\omega_c)\ll1$ and thus the resonant effects in this case are negligible.

\end{appendices}

\bibliographystyle{unsrt}
\bibliography{EfficientBalance}

\end{document}